\documentstyle[12pt]{article}
\begin{document}
\renewcommand{\baselinestretch}{1.5}
\begin{center}
{\bf  A NEW CANDIDATE FOR  THE DARK MATTER} 
\end{center}
\vspace{1in}
\begin{center}
{\bf AFSAR ABBAS} \\
{Institute of Physics} \\
{Bhubaneswar-751005, India} \\
{e-mail : afsar@iopb.ernet.in}
\end{center}
\vspace{1.5in}
\begin{center}
{\bf Abstract}
\end{center}

It is shown that if there are two distinct phase transition temperatures
to the hadronic phase ( as is currently favoured), then QCD 
demands a condensation to a weakly interacting massive glueball. 
This glueball then becomes a natural candidate for the 
cosmological dark matter. Note that this does not involve any
exotic extensions beyond the Standard Model of particle physics.
In this revised version at the end of the paper I have added a
new section  called ADDENDUM which gives new arguments as to
how this very heavy glueball can be stable and hence be a dark
matter candidate.
\vspace{1.5in}

\newpage
Among the few facts of cosmology is the existence of dark matter (DM) \cite{re1}.
In recent years it has become one of the most active areas of reseach
in physics and astronomy \cite{re2}. There is no dearth of DM candidate,
both baryonic and non-baryonic \cite{re3,re4}. The existence of none
of these has been satisfactorily demonstrated as of now \cite{re1,re2,re3,re4}.
It may be noted that most of these candidates are based on particle
physics models which are some kind of extensions beyond the Standard Model
(SM ) (of particle physics). All these are highly speculative as
these themselves have not been observed anywhere else \cite{re5}.

It is a common belief that the SM of particle physics based on the group
$ SU(3)_{c} \otimes SU(2)_{L} \otimes U(1)_{Y} $ is not rich enough
to justifiably accomodate the full complexity of the DM problem. 
(And hence the necessity to go beyond the SM, as stated above ).
In this paper I shall however show that actually there is new candidate
of DM - which is glueball, whose existence is predicted within the
framework of the SM itself.

Because of the non-abelian character of the $ SU(3)_{c} $ (on which the
Quantum Chromodynamics (QCD) is based) the gluons self interact. This
interaction does not require the existence of matter particles 
(i.e quarks etc). Hence in QCD on very general grounds one expects
the existence of the bound state of these gluons, the so called 
glueballs . Much work has been done on these glueballs both theoretically
and experimentally (see \cite{re6} for references) but no definitive
glueball candidate has been identified so far.

The currently very active ultrarelativistic heavy-ion collision programme
is being pursued with the hope of attaining the quark - gluon plasma
(QGP) phase of the hadronic matter. Earlier it was expected that
the hadron to QGP phase transition would take place at a single
temperature  $ T_{c}   \sim  150 - 200  $ MeV  \cite{re6}. 
Working in pure $ SU(3)_{c} $ theory
 (where no quarks were present) it was
shown that, free massless gluons which existed above $ T_{c} $ condense
through first order phase transition to a weakly interacting gas of
massive glueballs below $ T_{c} $ \cite{re7}. There has been discussion
as to whether the transition is first order or second order. Here for
the sake of simplicity we take the view that this phase transition is
first order \cite{re8}.

Recently it has been becoming popular to talk of two distinct phase
transition temperatures in QGP \cite{re9}. For example it has been
argued that the equilibrium of gluons takes place in time
 $ \tau_{g}   \sim   \frac{1}{2} $  fm/c while 
 the production and equilibriation of quarks
takes place in $ \tau_{q}   \sim  2 $   fm/c  .  Hence one has a hotter
pure gluon plasma. It has been shown \cite{re9} that the pure gluon plasma
transition temperature $ T_{g}  \sim   400 $ MeV  and for the quark phase
$ T_{q}   \sim   250  $ MeV .

We accept this two stage transition temperatures for QGP and turn the
argument around. In the big - bang scenario as the universe cools it
goes through various phase transitions \cite{re1,re2,re3,re4,re5}.
After the Electro - Weak symmetry is broken at temperature 
 $ \sim   200  $ GeV  , the particle content is $\gamma$,  $\nu_{e}$,
 $\bar{\nu_{e}}$,  $\nu_{\mu}$,  $\bar{\nu_{\mu}}$,  $\nu_{\tau}$,
 $\bar{\nu_{\tau}}$,  $e^{+}$ , $e^{-}$ ,  $\mu^{+}$ ,  $\mu^{-}$,
 $ u $,  $\bar{u}$,  $ d$,  $\bar{d}$,  and  gluons.\\ 
This remains so until the temperature 
cools down to  $ \sim  400  $ MeV . At this instant the first hadronic
phase transition takes place. Most ( or all ) gluons undergo first order
phase transition and condense to massive weakly interacting glueballs.
there is a freeze out and they decouple with the rest of the particles.
There are very few free gluons present immediately after this
phase transition. However very soon due to the reaction 
$ qq \rightarrow gg $  the quarks " dress " themselves up. 
By the time the temperature drops to $ \sim  250 $ MeV, there are 
enough quarks and gluons  present to lead to the normal QGP phase transition 
to the standard hadronic matter.

The weakly interacting massive glueballs become a relic species in the
universe. In the lowest order the glueball does not couple to photons
and hence would be an ideal candidate for  the dark matter. Hence there exists
within the framework of SM an object - the glueball which here is proposed
to be the new DM candidate.

To get an estimate of the mass of the glueball let us assume that when
it decouples it is non relativistic : 
\begin{displaymath}
 T_{g} << m_{g}  
 \end{displaymath}
Hence we obtain a conserved number N \cite{re4} 
\begin {equation}
N = \frac{n}{s} = 0.145 (\frac{g}{g_{*}}) ( \frac{m_{g}}{T_{g}})
\exp(-\frac{m_{g}}{T_{g}})
\end{equation}

Where $ g_{*} $ is the number of species in equlibrium when the glueball
decouples. Taking $ \rho_{c} $ to be $ 1.06 \times 10^{4} h^{2}
eV-cm^{-3} $ and imposing the constraint 
\begin{equation}
(\Omega  h^{2})_{glueball}  \le  1
\end{equation}
 
 If $ T_{g}  \sim  400  $ MeV   \cite{re9} then we find that 
$ m_{g} \ge 44.6 $ GeV.  So our estimate is that the glueball 
constituting the DM of the universe  is very heavy  $ \ge 44.6 $ GeV.

 All the estimates of the glueball studied theoretically \cite{re6,re7}
 have been of the order of 1 Gev for the lightest $ 0^{++} $ glueball.
 In all the extensive searches  for the glueball none has yet been detected.
 It's possible that we have not really understood the complete ( mainly
 non-perturbative ) dynamics of QCD \cite{re6,re7}. Neverthless the DM
 candidate glueball $ m_{g}  \ge  44.6 $ GeV  found here can not be the
 light glueball suggested by others. Clearly  work has to be
 done to understand how one can obtain such a massive glueball.
 Is it some long lived metastable state arising due to the complex
 dynamics of QCD ?  What kind of collectivity does it involve
 and what kind of phase transition  in QCD will give this ?

 In summary, if we accept the currently popular two distinct hadron phase
 transition temperatures in QGP \cite{re9}, then pure glueball condensation is
 naturally demanded by QCD. This then becomes a natural candidate for
 DM. This glueball is necessarily very massive too. The interesting point
 is  that this dark matter candidate is arising from the structure of the
 currently most successful model of particle physics -the Standard Model
 and does not involve any exotic extensions beyond it.

\newpage 

{\bf ADDENDUM}

So what is this heavy glueball and why is it so stable?
Note that we used the pure gluon plasma transition temperature
$ T_{g} \sim 400 $ MeV and for the quark phase $ T_{q} \sim 200 $
MeV. We find that a heavy glueball ($ m_{g} \ge 44.6 $ GeV)
arises due to the phase transition at $ \sim 400 $ MeV and the
rest of the hadron matter is created at $ \sim 200 $ MeV. There is
no a priori reason why the QCD phase transitions at these two
distinct temperatures have the same nature. The nature of the QCD
phase transition at 250 MeV , which leads to the standard hadrons has
been well studied in recent years and we can take it as rather well
understood [7]. So we do not mess around with this.
But what do we really know of the QCD phase transition at
$ \sim 400 $ MeV ? Is it not possible that this phase transition is
slightly different from the other one ? Hypothesysing this way
we can right away understand the nature of the heavy glueball
which has been proposed as a dark matter candidate.

Let us assume that the QCD phase transition at $ T_{g} \sim 400 $
MeV leads to the version of QCD as discussed by De Rujula, Giles
and Jaffe , Phys. Rev. {\bf D17} (1978) 285. Their framework is 
renormalizable spontaneously broken version of QCD. Color SU(3)
remains an exact global symmetry. The successful phenomenology
of color-singlet hadrons is altered very little and at the 
presently available energies the model is not inconsistent with
any experimental information. There is atleast one stable
hadron for every representation of color SU(3).Note that in our
picture the phase transition at $ T_g \sim 400 $ MeV leads to
only glueball condensation ( remember quark condensation takes 
place later at $ T_{q} \sim 250 $ MeV as per standard unbroken 
QCD ). Hence the glueballs that arise belong to the octet 
representation of SU(3). THey are heavy and stable.
This is the glueball dark matter candidate discussed above.


\end{document}